# Ballistic Transport in Metallic Nanotubes With Reliable Pd Ohmic Contacts

David Mann, Ali Javey, Jing Kong,[†] Qian Wang, and Hongjie Dai*

*Department of Chemistry and Laboratory for Advanced Materials,*

*Stanford University, Stanford, CA 94305, USA*

**ABSTRACT**

Contacting metallic single-walled carbon nanotubes by palladium (Pd) affords highly reproducible ohmic contacts and allows for detailed elucidation of ballistic transport in metallic nanotubes. The Pd ohmic contacts are more reliable than titanium (Ti) previously used for ballistic nanotube devices. In contrast, Pt contacts appear to give non-ohmic contacts to metallic nanotubes. For both ohmic and non-ohmic contacts, the length of the nanotube under the metal contact area is electrically turned off. Transport occurs from metal to nanotube at the edges of the contacts. Measurements with large numbers of Pd contacted nanotube samples reveal that the mean free path for defect scattering in SWNTs grown by chemical vapor deposition can be up to 4 μm. The mean free paths for acoustic phonon scattering are on the order of 500 nm at room temperature and » 4 μm at low temperatures.

* Corresponding author.  hdai@stanford.edu

[†] Current address: Department of NanoScience and DIMES, Delft University of Technology, Lorentzweg 1, 2628 CJ Delft, The Netherlands

Ohmic contacts with minimum contact resistance are important to the fundamental characterization and realization of high performance devices of electronic materials. Ohmic contacts to individual metallic single-walled carbon nanotubes (m-SWNTs) by Cr and Ti have enabled the observation of ballistic electron transport in SWNTs.[1,2] The hallmark of ballistic transport in SWNTs with ideal contacts is the conductance approaching the quantum limit $G=4e^2/h$ (resistance R~6.5 kΩ) and manifestation of phase coherent resonance transport at low temperatures.[1,2] More recently, ohmic contacts to semiconducting SWNTs (s-SWNTs) have been achieved with Pd electrodes.[3] The high work function of Pd and its favorable interactions with nanotube sidewalls affords largely suppressed Schottky barriers at the Pd/S-SWNT contacts, allowing for the observation of ballistic transport through the valence band of s-SWNTs. The on-states of ohmically contacted semiconducting SWNTs exhibits similar characteristics to ballistic metallic tubes with $G\sim 4e^2/h$ and current delivery capability reaching the optical phonon scattering limit of ~25 μA per tube.[3]

In this Letter, we report highly reliable ohmic contacts to metallic SWNTs made by Pd. Compared to other metals such as Ti, Pd is unique in giving ohmic contacts to m-SWNTs with very high reproducibility. In addition to two terminal m-SWNTs with ohmic contact, devices with a segment of the metallic tube covered by Pd or Pt are fabricated and investigated by transport measurements. Furthermore, the ohmically contacted m-SWNT devices allow for the elucidation of electron transport properties intrinsic to the nanotube material. Several important parameters for m-SWNTs grown by chemical vapor deposition (CVD) are estimated, including the mean free paths (mfp) for scattering by defects or imperfections and by acoustic phonons.

Individual metallic SWNT devices were fabricated by patterned CVD growth[4] of SWNTs on SiO$_2$/Si wafers, followed by electron beam lithography (EBL), metal deposition and liftoff to form source/drain (S/D) top-contacts.[5] The heavily doped Si substrate was used as back-gate and the thickness of the SiO$_2$ gate dielectric was either 500 nm or 67 nm. The thickness of the S/D metal (Pd or Ti) electrodes was ~30 nm deposited by electron beam evaporation in <10$^{-7}$ Torr vacuum. Devices with Pd S-D contacts were annealed at 220 °C in Ar for 10 minutes. In most cases, Pd ohmic contacts to m-SWNTs were found to form without the annealing step.

We first present results obtained with metallic SWNTs contacted by Ti at the S/D. Previously, we have reported ballistic transport in ohmically contacted relatively short (L~200 nm) m-SWNTs with Ti S/D electrodes.[2] Since then, we have observed ballistic transport in CVD grown metallic tubes with lengths up to 4 μm. Fig. 1a shows an ultra-straight nanotube (diameter d ~ 1.7 nm) with 4 μm length between Ti S/D contacts. The conductance of the device (measured under $V_{ds}$=1 mV) exhibits little gate-dependence as expected for metallic tubes, and monotonically increases from ~ $e^2/h$ (R~32 kΩ) to ~$3e^2/h$ (R~8.6 kΩ) as the sample is cooled from 290 K to 4 K. Between 4K and 300 mK in a $^3$He cryostat, pronounced slow conductance oscillations vs. gate voltage ($V_g$) is observed, and the differential conductance $dI/dV_{ds}$ vs. $V_g$ and $V_{ds}$ exhibits an interference pattern with conductance peaks and valleys at $V_{ds}$=0 (Fig. 1c). The results strongly point to a nearly ideal Ti/m-SWNT sample with highly transparent contacts (transmission probability at the two contacts $t_{Ti,1}$~ $t_{Ti,1}$~0.85 and G~ $t_{Ti,1} \times t_{Ti,2} \times 4e^2/h$ ~$3e^2/h$). There are no significant defects along the 4 μm length of the nanotube since no conductance degradation due to weak localization[2,6] is observed down to 300 mK. The somewhat irregular interference pattern does point to a non-ideal Fabry-Perot resonator.[1] This irregularity is attributed to minor disorder or inhomogeneity along the full length of the tube. A 3-fold conductance increase from 290 K to 4 K is observed and attributed to quenching of acoustic phonon (AP) scattering.[2,3,6,7] Thus for this particular device, one can glean that the mfp for defect scattering in the SWNT is $l_d > 4$ μm, and the mfps for acoustic phonon scattering are $l_{AP}(290K) < 4$ μm and $l_{AP}(300 mK) > 4$ μm.

The nearly ideal long ballistic SWNT sample with Ti contacts shown above is admittedly rare. It is also found that ohmic contacts to m-SWNTs by Ti has a limited success rate of ~ 10-20%. Relatively large fluctuations in contact resistance (10-100 kΩ) exist in different batches of devices. The ballistic mfp of 4 μm is observed once out of dozens of 3-4 μm long m-SWNTs.

In contrast to Ti, Pd affords ohmic contacts to metallic tubes with high reproducibility. Nearly all of the Pd contacted metallic SWNTs with lengths L≤ 1 μm exhibit conductance on the order of $2e^2/h$ (R ~ 10-20 kΩ) at 290 K and approach $4e^2/h$ at 4K (Fig. 2b for L=300 nm, Fig. 3c and Fig. 4c for L=1 μm). At low temperatures, the devices exhibit clear interference patterns in dI/$V_{ds}$ vs. $V_g$ and $V_{ds}$ (Fig. 2c). The interference patterns for various lengths SWNTs are consistent with the Fabry-Perot resonators described by Liang et al.[1] Whenever the phase shift acquired by electrons during a round trip in a SWNT reaches $2L \times eV_{ds}/\hbar v_F = 2\pi$ where $v_F = 8.1 \times 10^5$ m/s is the Fermi's velocity, a resonance peak occurs at $V_{ds}=V_c= \pi\hbar v_F/eL$. Plotting $V_c$ vs. 1/L gives a line with a slope[1] of $hv_F/2e$ = 1.670 mV·μm, which is the case in our Pd-contacted m-SWNT samples (Fig. 2d).

The mfp for defect scattering is a parameter specific to each individual SWNT and can be up to $l_d \sim 4$ μm (Fig. 1). Based on measurements of tens of Pd/m-SWNT devices, we conclude that the average mfp for defect scattering is $l_d \sim 1$ μm for our CVD grown nanotubes. For most devices with longer tube length, conductance vs. temperature data tend to show a downturn at low temperatures, signaling weak localization effects due to imperfections in the tube. It is also found that smaller diameter SWNTs (< 1.5 nm) are more likely to have various degrees of bends along the tube length. These tubes tend to become more insulating at low temperatures and have shorter $l_d$ due to imperfections related to mechanical deformation.

We observe that the conductance of shorter m-SWNTs exhibits weaker temperature dependence than longer ones. The L=4 μm long m-SWNT exhibits a 3-fold conductance increase (from $e^2/h$ to $3e^2/h$) from 290K to 4 K (Fig. 1b). On the other hand, the L~300 nm tubes typically exhibit a 30% conductance increase (from $2.5e^2/h$ to $3.2e^2/h$) upon cooling. The resistance in m-SWNTs due to acoustic phonon scattering can be written as,

$$R_{AP}(T) = (h/4e^2)[L/l_{AP}(T)]$$

Fitting of the measured resistance of long and short SWNTs to this expression (the physical meaning of which is that resistance due to acoustic phonon scattering scales with tube length L, and is negligible if $l_{AP} \ll L$) provides a rough estimate of $l_{AP}(290K) \sim$ 500 nm for acoustic phonon scattering at room temperature.

In Fig.3c, we show that a 1 μm long Pd contacted m-SWNT exhibits ballistic transport with conductance up to $3.3e^2/h$ at 1.5 K. On the same tube, a similar device is fabricated except for a Pd stripe covering a ~ 0.5 μm long segment of the tube between the Pd S/D (Fig. 3a&3b). Electrical measurements reveal that the conductance of the partially Pd-covered SWNT is limited below $2e^2/h$ (Fig. 3d; the uncovered device is limited below $4e^2/h$), and the interference pattern exhibits a characteristic energy scale ($V_c$~8.5 mV along the bias axis, Fig. 3e) approximately corresponding to SWNT resonators with tube length of L~200 nm which matches the lengths of the two uncoated tube segments. This suggests that the Pd stripe has divided the nanotube into two Fabry-Perot resonators in series. It is clear that the electron transmission probability at the four Pd-SWNT contact junctions is near unity ($t_{pd}$~0.9) and the nanotube segment covered by Pd is electrically turned off. This result illustrates the high reproducibility of Pd ohmic contacts to m-SWNTs. It also shows that electron transport into (or out of) an m-SWNT from (or into) a Pd top-contact occurs at the edge of the Pd electrode.

Different results are obtained with an m-SWNT partially covered with Pt along its length between Pd S/D (Fig. 4a&4b). At 290 K, the conductance measured between Pd S/D is 3 times lower (G~0.4$e^2/h$; R~65k$\Omega$; data not shown) than the same tube without Pt coverage (Fig. 4c, G~1.5$e^2/h$; R~17k$\Omega$;). Multi-probe measurements (with the Pt strip as an electrode) reveal that the resistance between the Pd S/D equals the sum of the resistance measured between S(Pd)-Pt and Pt-Pd(D). These results lead to the conclusion that, first, the Pt-SWNT contact is non-ohmic (non-ohmic contacts have been reported by Dekker et al. previously[8,9]) and has a resistance of ~ 20 k$\Omega$. Second, despite the relatively low transmission probability at Pt-SWNT contacts, electrons mostly transmit in and out of the Pt strip with the Pt covered section turned off. At 1.5 K, the SWNT without Pt coating exhibits a high conductance of ~ 3$e^2/h$ and interference effect (Fig. 4c). In strong contrast, the Pt coated device turns more insulating below ~ 50 K with a clear conductance dip in $dI/dV_{ds}$ vs. $V_{ds}$ at $V_{ds}=0$ (Fig. 4d).  The pronounced conductance dips (for some $V_g$) appear to be consistent with the known effect of Coulomb blockade in single-tunnel junction[10,11] devices (the Pt covered sample can be considered as two back-to-back Pd-SWNT-Pt systems each with an ohmic contact and a tunnel junction). Detailed results and analysis of this phenomenon will be presented elsewhere.

We have shown here that Pd gives highly reliable and reproducible ohmic contacts to metallic SWNTs, a result not surprising considering ohmic contacts made to semiconducting nanotubes by Pd.[3] Ti ohmic contacts to m-SWNT are less reliable despite excellent wetting of SWNT sidewalls by Ti.[12,13]  This could be due to the high chemical reactivity of Ti toward oxidation, and the quality of contact is sensitive to metal deposition conditions such as vacuum.  Pt contacts are non-ohmic to both metallic and semiconducting SWNTs.  The chemical nature of the Pd-SWNT and Pt-SWNT interfaces must be responsible for the differences in electrical transmissions, and remains elusive at the present time.  In the metal-on-top contact configuration, for both ohmic and non-ohmic contacts, the bulk length of the tube under the metal appears to be electrical off, and transport occurs from metal to nanotube at the edges of the contact.  This is consistent with the insensitivity of contact resistance to the contact length for SWNTs. With large numbers of ohmically contacted m-SWNT samples, we find that the mean free path for defect scattering in our CVD grown material is typically 1 µm and can be up to 4 µm.  The mean free paths for acoustic phonon scattering are on the order of 500 nm at room temperature, and > 4 µm below 4 K.

This work was supported by ABB Group Ltd, MARCO MSD Focus Center, DARPA/Moletronics, a Packard Fellowship, Sloan Fellowship and a Dreyfus Teacher-Scholar.

**Figure Captions**

**Figure 1.** (a) An atomic force microscopy (AFM) image of a 4 µm long SWNT (d ~ 1.7 nm) contacted by Ti at S/D. (b) Conductance vs. $V_g$ curves recorded at various temperatures from 290 K to 4K. (c) Differential conductance dI/dV vs. $V_g$ and $V_{ds}$ recorded at 300 mK. The arrow points to the characteristic conductance peak at $V_c$ ~ 0.75mV for the 4 µm long tube.

**Figure 2.** (a) An AFM image of a 300 nm SWNT (d ~ 3 nm) contacted by Pd electrodes. (b) G vs. $V_g$ curves from 290 to 4K. (c) dI/dV vs. $V_g$ and $V_{ds}$ recorded at 1.5K showing clear conductance oscillations with $V_c$=7 mV (pointed by an arrow). (d) Plot of $V_c$ vs. the inverse of tube length L. The slope of the line is $hv_F/2e$ = 1.670 mV.µm where $v_F$ = $8.1 \times 10^5$ m/s

**Figure 3.** (a) A schematic drawing of two types of devices formed on a single SWNT. The first type consists of a ~ 1 µm long SWNT between two Pd S/D electrodes. The second type is similar except for a Pd strip covering part of the SWNT between two Pd electrodes. All of the metal structures were formed by one step of EBL, Pd metal deposition and liftoff process. (b) An AFM image of the SWNT (d ~3 nm) structures depicted in (a). (c) G vs. $V_g$ recorded from 290 K to 4 K for the SWNT without Pd coverage between S/D. (d) G vs. $V_g$ recorded from 290 K to 4 K for the device with additional Pd coating in the middle. (e) dI/dV vs. $V_g$ and $V_{ds}$ at 1.5 K. Conductance peaks at $V_c$ = 8.5mV (at the arrow point) correspond to tube resonators ~200 nm in length (similar to the lengths of the tube sections free of Pd coating).

**Figure 4.** (a) A schematic drawing of two types of devices formed on a single SWNT. The first type consists of a ~ 1 μm long SWNT between two Pd S/D electrodes. The second type is similar except for a Pt electrode placed onto the SWNT between two Pd electrodes. The Pt electrode was formed by a second step of EBL, Pt deposition (by sputtering) and liftoff, after the first step for Pd. The offset of the Pt electrode from the center was due to misalignment. (b) An AFM image of the SWNT structure (d ~ 3.2 nm) depicted in (a). (c) G vs. $V_g$ curves recorded from 290 K to 1.5 K for a SWNT between two Pd S/D contacts. The small, rapid conductance oscillations from 290 K - 50 K are attributed to noise in the measurement system. (d) $dI/dV_{ds}$ vs. $V_{ds}$ measured at 1.5 K (under two gate voltages) for the SWNT between two Pd contacts with an additional Pt electrode placed in between. The dip structure exists over the entire $V_g$ range scanned.

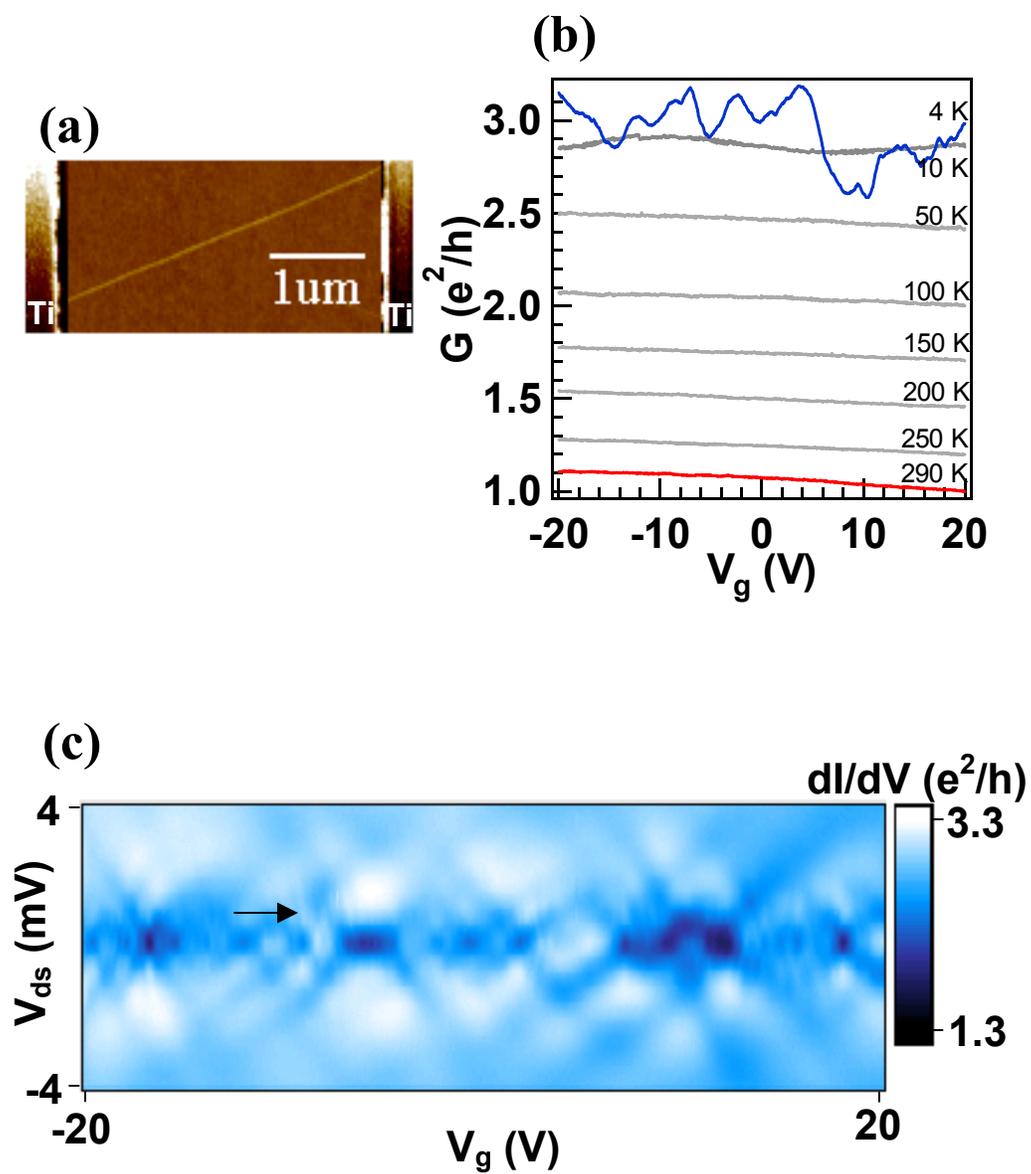

Figure 1

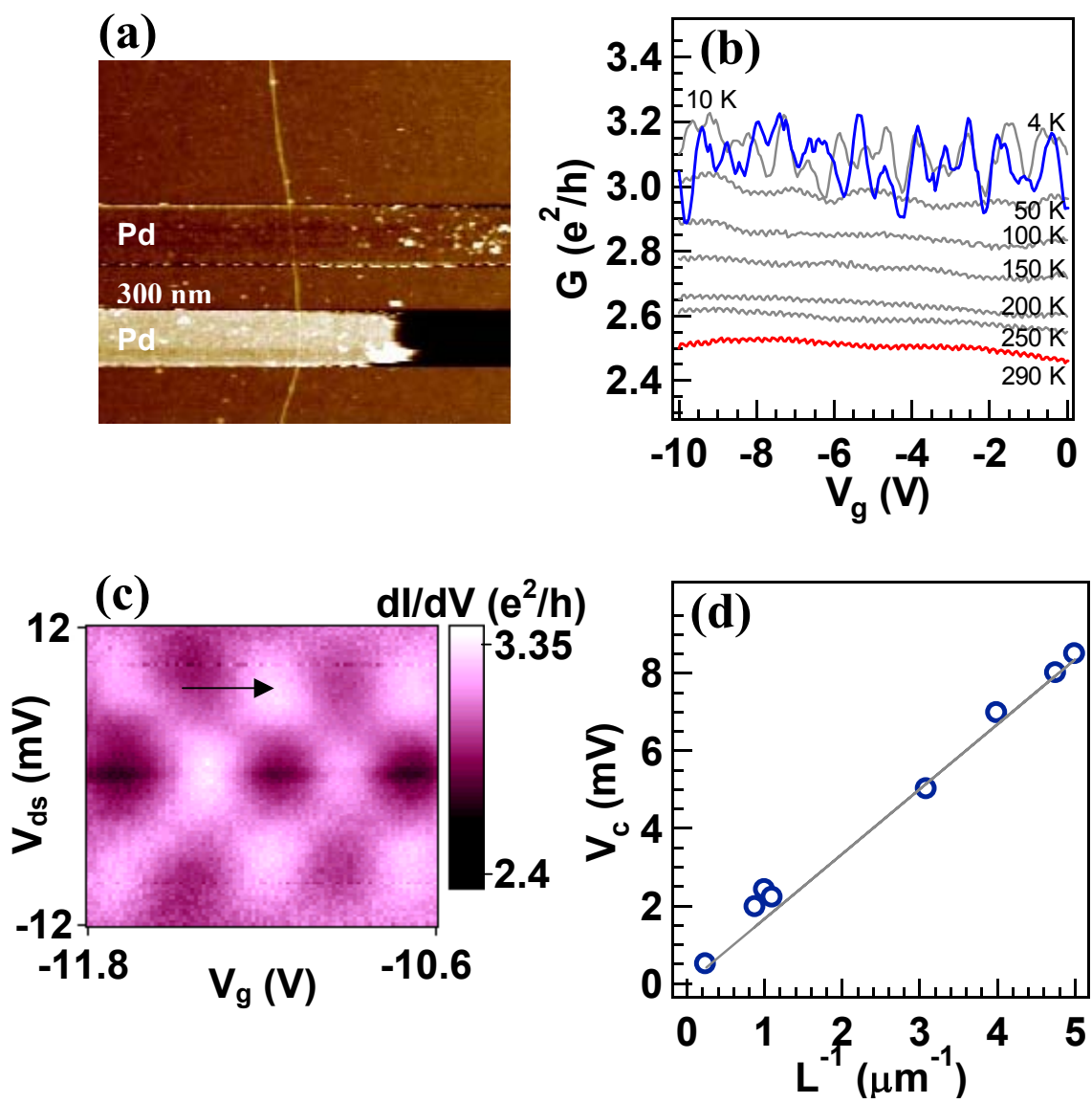
Figure 2

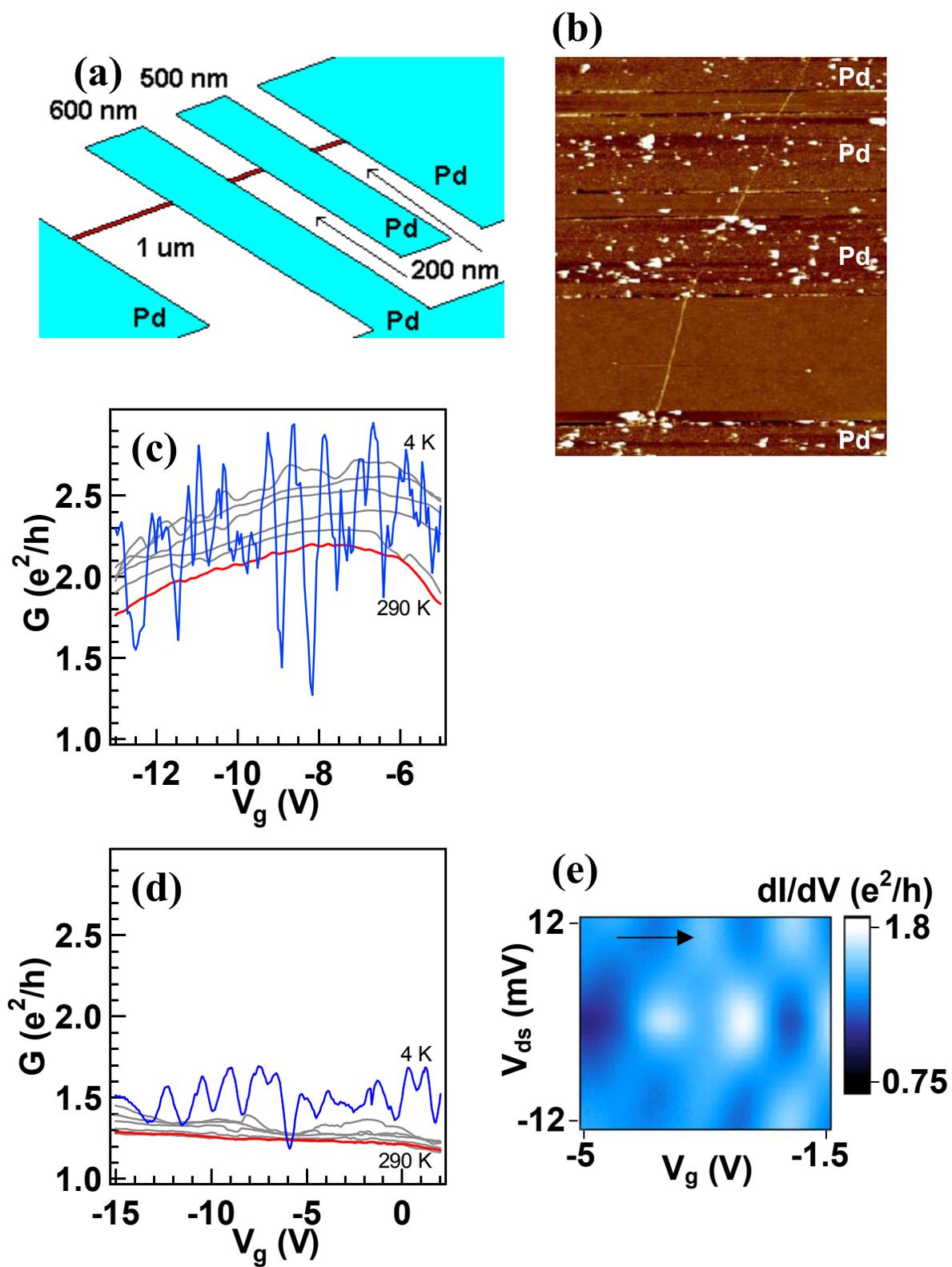

Figure 3

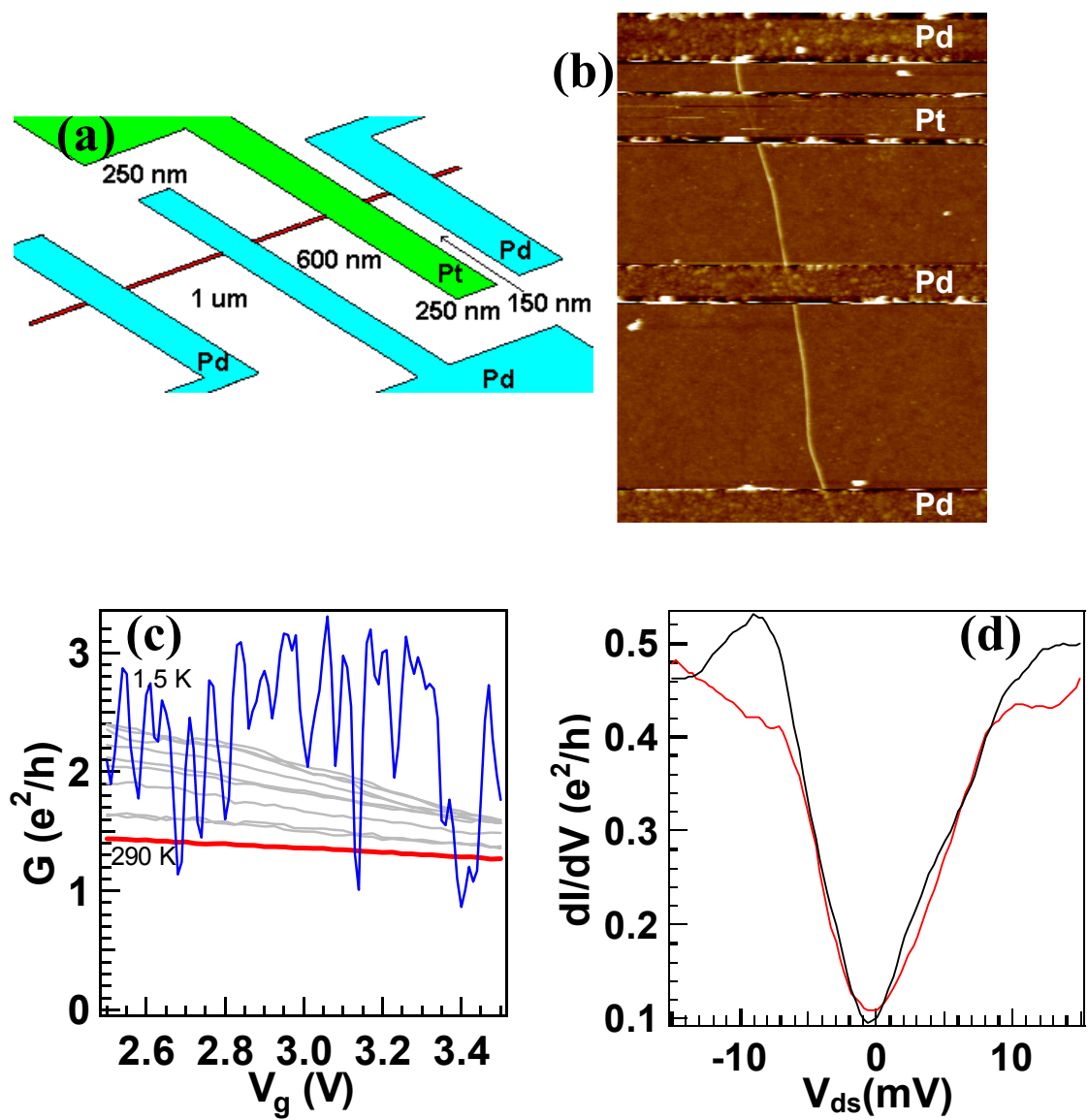

Figure 4